# Guessing probability under unlimited known-plaintext attack on secret keys for Y00 quantum stream cipher by quantum multiple hypotheses testing

**Takehisa Iwakoshi***
Quantum ICT Research Institute, Tamagawa University, 6-1-1, Tamagawa-Gakuen, Machida, Tokyo, Japan

**Abstract**. Although quantum key distribution is regarded as promising secure communication, security of Y00 protocol proposed by Yuen in 2000 for the affinity to conventional optical communication is not well-understood yet; its security has been evaluated only by the eavesdropper's error probabilities of detecting individual signals or masking size, the number of hidden signal levels under quantum and classical noise. Our study is the first challenge of evaluating the guessing probabilities on shared secret keys for pseudorandom number generators in a simplified Y00 communication system based on quantum multiple hypotheses testing theory. The result is that even unlimitedly long known-plaintext attack only lets the eavesdropper guess the shared secret keys of limited lengths with a probability strictly <1. This study will give some insights for detailed future works on this quantum communication protocol.

***Takehisa Iwakoshi, E-mail: t.iwakoshi@lab.tamagawa.ac.jp**

## 1 Introduction

Recent heating-up in the development race of Quantum Computers brings attention to secure communications which are resistant to Quantum Computers. Quantum Key Distribution (QKD) has been said to be the most promising technology to protect communications from cryptanalysis even with Quantum Computers.

On the other hand, Y00 protocol proposed by H. P. Yuen (its original name is αη) in 2000 [1–3] is compatible to conventional high-speed and long-distance optical communication technologies while it sends messages directly and hides them under quantum noise enhancing the security of conventional cryptographies [4–14]. However, its security has not been well-understood except some evaluations by unicity distance [15, 16], numerical analyses by masking sizes that are the numbers of signal levels hidden under quantum noise [17, 18], or error probability in eavesdropping on individual signals [19]. Therefore, since Fast Correlation Attack on Y00 protocol was found [20, 21], Y00 protocol has been believed to be computationally secure while QKDs are Information-Theoretically secure, although a countermeasure “Irregular Mapping” was equipped on Y00 systems as a countermeasure to Fast Correlation Attack [22].

This study gives the first example of security analysis on Y00 protocol with unlimitedly long Known-Plaintext Attack (KPA) by Quantum Multiple Hypotheses Testing Theory. It shows that Y00 protocol is secure with guessing probabilities on the two shared secret keys of 128 bits strictly less than one even under unlimitedly long KPA. However, as time passes the attack increases the guessing probability, hence fresh keys have to be sent instead of messages before Y00 communication systems are breached. In this case, if the probability distribution of the provided fresh key is uniform, the guessing probability on the key has to be evaluated by Ciphertext-Only Attack (COA). There are still some assumptions in this study, therefore this study is not rigorous yet. However, it will give a better understanding of the security of Y00 protocol, and some insights for detailed future works.

One of the major methods to evaluate Information-Theoretic security has been to calculate the guessing probability on secret keys since C. E. Shannon [23] and in the literature following [24, 25]. Therefore, this study follows these concepts as well to evaluate the security of Y00 protocol.

## 2 Known works on security evaluations on Information-Theoretic Secure Cryptography

The founder of the Information Theory, C. E. Shannon proved that there is that “perfect secrecy” is satisfied only when the length of the encryption key $\boldsymbol{k}$ with its probability distribution Independent and Identically-Distributed (IID) has to be longer than the plaintext $\boldsymbol{x}$, which is One-Time Pad (OTP)[23]. Then the ciphertext string $\boldsymbol{c}$ is given by the modulo-2 addition of $\boldsymbol{x}$ and $\boldsymbol{k}$.

$$\boldsymbol{x} + \boldsymbol{k} = \boldsymbol{c} \pmod 2 . \tag{1}$$

Therefore,

$$\Pr(\boldsymbol{x} \mid \boldsymbol{c}) = \Pr(\boldsymbol{x}). \tag{2}$$

In 2012, M. Alimomeni and R. Safavi-Naini proposed "Guessing Secrecy," generalizing Shannon's concept [24]; the encryption is not perfectly secure, but the key is obtained only probabilistically by Eve's guess as

$$\sum_{c \in C} \Pr(\boldsymbol{c}) \max_{x \in X} \Pr(\boldsymbol{x} \mid \boldsymbol{c}) = \max_{x \in X} \Pr(\boldsymbol{x}). \tag{3}$$

In 2013, M. Iwamoto and J. Shikata proposed "Worst-case Guessing Secrecy" [25], considering the worst scenario such as

$$\max_{(x,c) \in (X,C)} \Pr(\boldsymbol{x} \mid \boldsymbol{c}) = \max_{x \in X} \Pr(\boldsymbol{x}). \tag{4}$$

Therefore, this study follows the above concepts "guessing probability on the key" to evaluate the security of Y00 protocol.

## 3 Security of Conventional Stream Ciphers under Long Known-Plaintext Attack

This section treats the security of conventional stream ciphers with KPA; those are not randomized by quantum noise to give a better understanding on the security of Y00 protocol which is a stream cipher randomized by quantum noise. In conventional stream ciphers, a shared secret key $\boldsymbol{k}$ is fed into the Pseudo-Random-Number Generator (PRNG) to generate a key stream $\boldsymbol{s}$. A plaintext string $\boldsymbol{x}$ from a sender, Alice, is converted into a ciphertext string $\boldsymbol{c} = \boldsymbol{x} + \boldsymbol{s} \bmod 2$. To decode $\boldsymbol{c}$, the receiver, Bob, feeds the shared $\boldsymbol{k}$ into his common PRNG, then recovers $\boldsymbol{x} = \boldsymbol{c} + \boldsymbol{s} \bmod 2$. If Eve has the same PRNG and knows $\boldsymbol{x}$ during a period of $\boldsymbol{s}$, she obtains $\boldsymbol{s}$ completely, hence her correspondence table of $\boldsymbol{k} \leftrightarrow \boldsymbol{s}$ recovers the original key $\boldsymbol{k}$ no matter how much computationally complex the key expansion process is. Then, Eve can read all messages from the next period. In terms of conditional probability, this means,

$$\Pr(\boldsymbol{s} \mid \boldsymbol{c}, \boldsymbol{x}) = \Pr(\boldsymbol{k} \mid \boldsymbol{c}, \boldsymbol{x}) = 1 . \tag{5}$$

## 4 Security of Y00 under KPA during Least Common Multiple of PRNGs' Periods

This section describes the security of Y00 protocol with KPA during the Least Common Multiple (LCM) of the two PRNGs' periods using Quantum Multiple Hypotheses Testing Theory [26, 27].

### *4.1 Principles of Binary Y00 Quantum Stream Cipher*

To start Y00 protocol, the legitimate users Alice and Bob have to share a secret key $\boldsymbol{k}$. Then they expand $\boldsymbol{k}$ into a key stream $\boldsymbol{s}$ using a common PRNG equipped in each transmitter/receiver. Then $\boldsymbol{s}$ is chopped every $\log_2 M$ bits to form $M$-ary string $\boldsymbol{s}(t)$ of time slot $t$, while a message bit $x(t)$ is encoded into a coherent state $|\alpha[m(t)]\rangle$ using $\boldsymbol{s}(t)$ as

$$m(t) := \mathrm{Map}\left[\boldsymbol{s}(t)\right] + M\left(\mathrm{Map}\left[\boldsymbol{s}(t)\right] + x(t) \bmod 2\right) . \tag{6}$$

Map[$\boldsymbol{s}(t)$] is a projection from $\boldsymbol{s}(t)$ to Map[$\boldsymbol{s}(t)$] ∈ {0, 1, 2, 3, … , $M-1$}. Therefore, the message bit $x(t) \in \{0, 1\}$ corresponds to a set of quantum states $\{|\alpha[m(t)]\rangle, |\alpha[m(t)+M]\rangle\}$ for even number Map[$\boldsymbol{s}(t)$], otherwise $\{|\alpha[m(t)+M]\rangle, |\alpha[m(t)]\rangle\}$. On the other hand, Bob's receiver sets an optimal threshold(s) to discriminate the set of quantum states. Therefore he decodes $x(t)$ since he knows Map[$\boldsymbol{s}(t)$] thanks to the common PRNG and the shared $\boldsymbol{k}$. On the other hand, the eavesdropper, Eve, has to discriminate $2M$-ary signals hidden under overlapping quantum and classical noise since she does not know whether Map[$\boldsymbol{s}(t)$] is even or odd, hence $x(t)$ neither.

When Eve launches Known-Plaintext Attack (KPA), a number of the hidden signal level under noise effectively halves, hence it might help Eve to guess $\boldsymbol{k}$.

To avoid the situation, Overlap-Selection-Keying (OSK) was proposed [28]. An additional common PRNGs with another shared key $\mathbf{\Delta k}$ are equipped in both a transmitter and a receiver to randomize the plaintext $\boldsymbol{x}$ with pseudo-random number $\Delta\boldsymbol{x}$ as

$$m(t) := \mathrm{Map}\left[\boldsymbol{s}(t)\right] + M\left(\mathrm{Map}\left[\boldsymbol{s}(t)\right] + x(t) + \Delta x(t) \bmod 2\right). \tag{7}$$

Then the transmitter Alice sends a coherent state $\rho(m(t))$ with classical randomizations named DSR and DER [19] although these are omitted in this study for simplicity.

Eve obtains coherent states separated from a beam-splitter $\rho'(m(t))$ and stores its time sequence in her quantum memory. Denote the quantum sequence $\rho'(\boldsymbol{x}, \boldsymbol{s}, \Delta\boldsymbol{x})$ with the splitting ratio $\eta$ as

$$\rho'\left(\boldsymbol{x}, \boldsymbol{s}, \Delta\boldsymbol{x}\right) := \left|\eta\alpha\left(\boldsymbol{x}, \boldsymbol{s}, \Delta\boldsymbol{x}\right)\right\rangle\left\langle\eta\alpha\left(\boldsymbol{x}, \boldsymbol{s}, \Delta\boldsymbol{x}\right)\right| = \bigotimes_{t=0}^{T-1} \left|\eta\alpha[m(t)]\right\rangle\left\langle\eta\alpha[m(t)]\right|. \tag{8}$$

Note that a set of $(\boldsymbol{s}, \Delta\boldsymbol{x}) \in (\boldsymbol{S}, \Delta\boldsymbol{X})$ is generated from $(\boldsymbol{k}, \mathbf{\Delta k}) \in (\boldsymbol{K}, \mathbf{\Delta K})$. Therefore there are only $2^{|\boldsymbol{K}|+|\Delta\boldsymbol{K}|}$ patterns of single sequences, although the number of single levels is $2M$ and the period of KPA is $T$. Hence, what Eve needs is not $2M\cdot T$-ary quantum decision theory but $2^{|\boldsymbol{K}|+|\Delta\boldsymbol{K}|}$ -ary one, no matter how long the key-stream lengths of $\boldsymbol{s}$ and $\Delta\boldsymbol{x}$ are. Therefore the main problem is whether Eve can determine the correct $(\boldsymbol{s}, \Delta\boldsymbol{x})$ in the LCM of the periods of $(\boldsymbol{s}, \Delta\boldsymbol{x})$, denoted as $T_{\mathrm{LCM}}$, like in case of the conventional stream cipher explained in Sec.3 or she needs longer than $T_{\mathrm{LCM}}$.

### *4.2 Brief Description of M-ary Quantum Detection Theory*

Before this section starts, here are some assumptions to be satisfied.

- The projection Map[·] stays unchanged during the running of Y00 protocol.
- Map[·] is known to Eve according to the Kerckhoffs' principle, as known as Shannon's Maxim.
- Map[·] is well-designed irregular mapping so that quantum noise covers all bits in $\boldsymbol{s}(t)$ equally.

The set of Eve's measurement operators $\{E(\boldsymbol{s}, \Delta\boldsymbol{x} \mid \boldsymbol{x})\}$ satisfies

$$\sum\nolimits_{(\boldsymbol{s},\Delta\boldsymbol{x})\in(\boldsymbol{S},\Delta\boldsymbol{X})} E\left(\boldsymbol{s}, \Delta\boldsymbol{x} \mid \boldsymbol{x}\right) = I\ . \tag{9}$$

By Born rule, the measurement operator $E(\boldsymbol{s}', \Delta\boldsymbol{x}' \mid \boldsymbol{x})$ gives Eve a measurement result $(\boldsymbol{s}', \Delta\boldsymbol{x}') \in (\boldsymbol{S}, \boldsymbol{\Delta X})$ from a quantum state $\rho'(\boldsymbol{x}, \boldsymbol{s}, \Delta\boldsymbol{x})$ with a probability of

$$\mathrm{tr}\left[E\left(\boldsymbol{s}',\Delta\boldsymbol{x}' \mid \boldsymbol{x}\right)\rho'\left(\boldsymbol{x},\boldsymbol{s},\Delta\boldsymbol{x}\right)\right]=\Pr\left(\boldsymbol{s}',\Delta\boldsymbol{x}' \mid \boldsymbol{x},\boldsymbol{s},\Delta\boldsymbol{x}\right). \tag{10}$$

Quantum multi-hypotheses testing theory based on the Bayes criterion is applicable to decide which $(\boldsymbol{s}', \Delta\boldsymbol{x}')$ is the most possible. Let the Bayes cost in the theory be as described in Eq.(11).

$$C\left(\boldsymbol{s},\Delta\boldsymbol{x},\boldsymbol{s}',\Delta\boldsymbol{x}' \mid \boldsymbol{x}\right):=-\delta_{\boldsymbol{s},\boldsymbol{s}'}\delta_{\Delta\boldsymbol{x},\Delta\boldsymbol{x}'}. \tag{11}$$

When the prior probability is $\Pr(\boldsymbol{s}, \Delta\boldsymbol{x})$, the average Bayes cost is

$$\mathrm{Ex}\left[C\right]=-\sum\nolimits_{(\boldsymbol{s},\Delta\boldsymbol{x}),(\boldsymbol{s}',\Delta\boldsymbol{x}')\in(\boldsymbol{S},\Delta\boldsymbol{X})}\Pr\left(\boldsymbol{s},\Delta\boldsymbol{x}\right)\delta_{\boldsymbol{s},\boldsymbol{s}'}\delta_{\Delta\boldsymbol{x},\Delta\boldsymbol{x}'}\,\mathrm{tr}\left[\rho'\left(\boldsymbol{x},\boldsymbol{s},\Delta\boldsymbol{x}\right)E\left(\boldsymbol{s}',\Delta\boldsymbol{x}' \mid \boldsymbol{x}\right)\right]. \tag{12}$$

The Hermitian risk operators are

$$\begin{aligned}W\left(\boldsymbol{x},\boldsymbol{s}',\Delta\boldsymbol{x}'\right)&:=\sum\nolimits_{(\boldsymbol{s},\Delta\boldsymbol{x})\in(\boldsymbol{S},\Delta\boldsymbol{X})}\Pr\left(\boldsymbol{s},\Delta\boldsymbol{x}\right)\left(-\delta_{\boldsymbol{s},\boldsymbol{s}'}\delta_{\Delta\boldsymbol{x},\Delta\boldsymbol{x}'}\right)\rho'\left(\boldsymbol{x},\boldsymbol{s},\Delta\boldsymbol{x}\right)\\&=-\Pr\left(\boldsymbol{s}',\Delta\boldsymbol{x}'\right)\rho'\left(\boldsymbol{x},\boldsymbol{s}',\Delta\boldsymbol{x}'\right)=-\Pr\left(\boldsymbol{s}',\Delta\boldsymbol{x}'\right)\left|\eta\alpha\left(\boldsymbol{x},\boldsymbol{s}',\Delta\boldsymbol{x}'\right)\right\rangle\left\langle\eta\alpha\left(\boldsymbol{x},\boldsymbol{s}',\Delta\boldsymbol{x}'\right)\right|\end{aligned}. \tag{13}$$

To minimize Eve's error probability, the necessary-and-sufficient conditions are [26]

$$\left[W\left(\boldsymbol{x},\boldsymbol{s},\Delta\boldsymbol{x}\right)-\Gamma\right]E\left(\boldsymbol{s},\Delta\boldsymbol{x} \mid \boldsymbol{x}\right)=E\left(\boldsymbol{s},\Delta\boldsymbol{x} \mid \boldsymbol{x}\right)\left[W\left(\boldsymbol{x},\boldsymbol{s},\Delta\boldsymbol{x}\right)-\Gamma\right]=0. \tag{14}$$

$$E\left(\boldsymbol{s},\Delta\boldsymbol{x} \mid \boldsymbol{x}\right)\left[W\left(\boldsymbol{x},\boldsymbol{s}',\Delta\boldsymbol{x}'\right)-W\left(\boldsymbol{x},\boldsymbol{s},\Delta\boldsymbol{x}\right)\right]E\left(\boldsymbol{s}',\Delta\boldsymbol{x}' \mid \boldsymbol{x}\right)=\boldsymbol{0}. \tag{15}$$

$$W\left(\boldsymbol{x},\boldsymbol{s},\Delta\boldsymbol{x}\right)-\Gamma\geq 0. \tag{16}$$

$$\Gamma:=\sum\nolimits_{(\boldsymbol{s},\Delta\boldsymbol{x})\in(\boldsymbol{S},\Delta\boldsymbol{X})}E\left(\boldsymbol{s},\Delta\boldsymbol{x} \mid \boldsymbol{x}\right)W\left(\boldsymbol{x},\boldsymbol{s},\Delta\boldsymbol{x}\right)=\sum\nolimits_{(\boldsymbol{s},\Delta\boldsymbol{x})\in(\boldsymbol{S},\Delta\boldsymbol{X})}W\left(\boldsymbol{x},\boldsymbol{s},\Delta\boldsymbol{x}\right)E\left(\boldsymbol{s},\Delta\boldsymbol{x} \mid \boldsymbol{x}\right). \tag{17}$$

Then Eve's maximum success probability of obtaining the correct $(\boldsymbol{s}, \Delta\boldsymbol{x})$ is,

$$\begin{aligned}\Pr\left(\boldsymbol{s},\Delta\boldsymbol{x} \mid \boldsymbol{x},\boldsymbol{s},\Delta\boldsymbol{x}\right)&=1-\left(1+\mathrm{tr}\,\Gamma\right)=-\mathrm{tr}\,\Gamma\\&=\sum\nolimits_{(\boldsymbol{s},\Delta\boldsymbol{x})\in(\boldsymbol{S},\Delta\boldsymbol{X})}\Pr\left(\boldsymbol{s},\Delta\boldsymbol{x}\right)\left\langle\eta\alpha\left(\boldsymbol{x},\boldsymbol{s},\Delta\boldsymbol{x}\right)\right|E\left(\boldsymbol{s},\Delta\boldsymbol{x} \mid \boldsymbol{x}\right)\left|\eta\alpha\left(\boldsymbol{x},\boldsymbol{s},\Delta\boldsymbol{x}\right)\right\rangle\end{aligned}. \tag{18}$$

Now, denote $E(\boldsymbol{s}, \Delta\boldsymbol{x} \mid \boldsymbol{x})$ as

$$E\left(\boldsymbol{s},\Delta\boldsymbol{x} \mid \boldsymbol{x}\right):=\left|\left(\boldsymbol{s},\Delta\boldsymbol{x} \mid \boldsymbol{x}\right)\right\rangle\left\langle\left(\boldsymbol{s},\Delta\boldsymbol{x} \mid \boldsymbol{x}\right)\right|. \tag{19}$$

From Eq.(15),

$$\begin{aligned}&\Pr(\boldsymbol{s},\Delta\boldsymbol{x})\langle(\boldsymbol{s},\Delta\boldsymbol{x}\mid\boldsymbol{x})|\eta\alpha(\boldsymbol{x},\boldsymbol{s}',\Delta\boldsymbol{x}')\rangle\langle\eta\alpha(\boldsymbol{x},\boldsymbol{s}',\Delta\boldsymbol{x}')|(\boldsymbol{s}',\Delta\boldsymbol{x}'\mid\boldsymbol{x})\rangle\\&=\Pr(\boldsymbol{s}',\Delta\boldsymbol{x}')\langle(\boldsymbol{s},\Delta\boldsymbol{x}\mid\boldsymbol{x})|\eta\alpha(\boldsymbol{x},\boldsymbol{s},\Delta\boldsymbol{x})\rangle\langle\eta\alpha(\boldsymbol{x},\boldsymbol{s},\Delta\boldsymbol{x})|(\boldsymbol{s}',\Delta\boldsymbol{x}'\mid\boldsymbol{x})\rangle\end{aligned}. \tag{20}$$

For pure states, from Eq.(8),

$$\langle\eta\alpha(\boldsymbol{x},\boldsymbol{s}',\Delta\boldsymbol{x}')|\eta\alpha(\boldsymbol{x},\boldsymbol{s}',\Delta\boldsymbol{x}')\rangle=1=\sum_{(\boldsymbol{s},\Delta\boldsymbol{x})\in(\boldsymbol{S},\Delta\boldsymbol{X})}\left|\langle\eta\alpha(\boldsymbol{x},\boldsymbol{s}',\Delta\boldsymbol{x}')|(\boldsymbol{s},\Delta\boldsymbol{x}\mid\boldsymbol{x})\rangle\right|^2. \tag{21}$$

Therefore, Eq.(20) gives $2^{2|\boldsymbol{K}|+2|\Delta\boldsymbol{K}|}-2^{|\boldsymbol{K}|+|\Delta\boldsymbol{K}|}$ equalities and Eq.(21) gives $2^{|\boldsymbol{K}|+|\Delta\boldsymbol{K}|}$ equalities. Thus, there are $2^{2|\boldsymbol{K}|+2|\Delta\boldsymbol{K}|}$ equations in total, while there are $2^{3|\boldsymbol{K}|+3|\Delta\boldsymbol{K}|}$ variables including $\{\Pr(\boldsymbol{s},\Delta\boldsymbol{x})\}$.

To remove remained variables $\{\Pr(\boldsymbol{s},\Delta\boldsymbol{x})\}$, apply Cauchy–Schwarz inequality to Eq.(18).

$$\begin{aligned}-\operatorname{tr}\Gamma&=\sum_{(\boldsymbol{s},\Delta\boldsymbol{x})\in(\boldsymbol{S},\Delta\boldsymbol{X})}\Pr(\boldsymbol{s},\Delta\boldsymbol{x})\left|\langle\eta\alpha(\boldsymbol{x},\boldsymbol{s},\Delta\boldsymbol{x})|(\boldsymbol{s},\Delta\boldsymbol{x}\mid\boldsymbol{x})\rangle\right|^2\\&\le\left[\sum_{(\boldsymbol{s},\Delta\boldsymbol{x})\in(\boldsymbol{S},\Delta\boldsymbol{X})}\Pr(\boldsymbol{s},\Delta\boldsymbol{x})^2\right]^{1/2}\left[\sum_{(\boldsymbol{s},\Delta\boldsymbol{x})\in(\boldsymbol{S},\Delta\boldsymbol{X})}\left|\langle\eta\alpha(\boldsymbol{x},\boldsymbol{s},\Delta\boldsymbol{x})|(\boldsymbol{s},\Delta\boldsymbol{x}\mid\boldsymbol{x})\rangle\right|^4\right]^{1/2}\end{aligned}. \tag{22}$$

Let Eve know the prior probability $\Pr(\boldsymbol{s},\Delta\boldsymbol{x})$ under Shannon's Maxim. Then Eve can choose her $\{E(\boldsymbol{s},\Delta\boldsymbol{x}\mid\boldsymbol{x})\}$ so that the equality of Eq.(23) is satisfied.

$$\Pr(\boldsymbol{s},\Delta\boldsymbol{x})=\frac{\left|\langle\eta\alpha(\boldsymbol{x},\boldsymbol{s},\Delta\boldsymbol{x})|(\boldsymbol{s},\Delta\boldsymbol{x}\mid\boldsymbol{x})\rangle\right|^2}{\sum_{(\boldsymbol{s},\Delta\boldsymbol{x})\in(\boldsymbol{S},\Delta\boldsymbol{X})}\left|\langle\eta\alpha(\boldsymbol{x},\boldsymbol{s},\Delta\boldsymbol{x})|(\boldsymbol{s},\Delta\boldsymbol{x}\mid\boldsymbol{x})\rangle\right|^2}. \tag{23}$$

Therefore the prior probability distribution $\{\Pr(\boldsymbol{s},\Delta\boldsymbol{x})\}$ vanishes as follows.

$$\max[-\operatorname{tr}\Gamma]=\frac{\sum_{(\boldsymbol{s},\Delta\boldsymbol{x})\in(\boldsymbol{S},\Delta\boldsymbol{X})}\left|\langle\eta\alpha(\boldsymbol{x},\boldsymbol{s},\Delta\boldsymbol{x})|(\boldsymbol{s},\Delta\boldsymbol{x}\mid\boldsymbol{x})\rangle\right|^4}{\sum_{(\boldsymbol{s},\Delta\boldsymbol{x})\in(\boldsymbol{S},\Delta\boldsymbol{X})}\left|\langle\eta\alpha(\boldsymbol{x},\boldsymbol{s},\Delta\boldsymbol{x})|(\boldsymbol{s},\Delta\boldsymbol{x}\mid\boldsymbol{x})\rangle\right|^2}. \tag{24}$$

The condition Eq.(23) satisfies Eq.(14) trivially, and Eqs.(15, 16) are converted as follows.

$$\begin{aligned}&\left|\langle\eta\alpha(\boldsymbol{x},\boldsymbol{s}',\Delta\boldsymbol{x}')|(\boldsymbol{s}',\Delta\boldsymbol{x}'\mid\boldsymbol{x})\rangle\right|^2\langle(\boldsymbol{s},\Delta\boldsymbol{x}\mid\boldsymbol{x})|\eta\alpha(\boldsymbol{x},\boldsymbol{s}',\Delta\boldsymbol{x}')\rangle\langle\eta\alpha(\boldsymbol{x},\boldsymbol{s}',\Delta\boldsymbol{x}')|(\boldsymbol{s}',\Delta\boldsymbol{x}'\mid\boldsymbol{x})\rangle\\&=\left|\langle\eta\alpha(\boldsymbol{x},\boldsymbol{s},\Delta\boldsymbol{x})|(\boldsymbol{s},\Delta\boldsymbol{x}\mid\boldsymbol{x})\rangle\right|^2\langle(\boldsymbol{s},\Delta\boldsymbol{x}\mid\boldsymbol{x})|\eta\alpha(\boldsymbol{x},\boldsymbol{s},\Delta\boldsymbol{x})\rangle\langle\eta\alpha(\boldsymbol{x},\boldsymbol{s},\Delta\boldsymbol{x})|(\boldsymbol{s}',\Delta\boldsymbol{x}'\mid\boldsymbol{x})\rangle\end{aligned}. \tag{25}$$

$$\begin{aligned}&\langle(\boldsymbol{s},\Delta\boldsymbol{x}\mid\boldsymbol{x})|\left[W(\boldsymbol{x},\boldsymbol{s},\Delta\boldsymbol{x})-\Gamma\right]|(\boldsymbol{s},\Delta\boldsymbol{x}\mid\boldsymbol{x})\rangle\sum_{(\boldsymbol{s},\Delta\boldsymbol{x})\in(\boldsymbol{S},\Delta\boldsymbol{X})}\left|\langle\eta\alpha(\boldsymbol{x},\boldsymbol{s},\Delta\boldsymbol{x})|(\boldsymbol{s},\Delta\boldsymbol{x}\mid\boldsymbol{x})\rangle\right|^2\\&=-\left|\langle\eta\alpha(\boldsymbol{x},\boldsymbol{s},\Delta\boldsymbol{x})|(\boldsymbol{s},\Delta\boldsymbol{x}\mid\boldsymbol{x})\rangle\right|^4+\sum_{(\boldsymbol{s},\Delta\boldsymbol{x})\in(\boldsymbol{S},\Delta\boldsymbol{X})}\left|\langle\eta\alpha(\boldsymbol{x},\boldsymbol{s},\Delta\boldsymbol{x})|(\boldsymbol{s},\Delta\boldsymbol{x}\mid\boldsymbol{x})\rangle\right|^4\ge0\end{aligned}. \tag{26}$$

Therefore, Eq.(26) originated from Eq.(16) is also satisfied while a new condition is Eq.(25). The absolute value of Eq.(25) is

$$\begin{aligned}&\left|\left\langle \eta\alpha(\boldsymbol{x},\boldsymbol{s}',\Delta\boldsymbol{x}')\middle|(\boldsymbol{s}',\Delta\boldsymbol{x}'\,|\,\boldsymbol{x})\right\rangle\right|^4\left|\left\langle (\boldsymbol{s},\Delta\boldsymbol{x}\,|\,\boldsymbol{x})\middle|\eta\alpha(\boldsymbol{x},\boldsymbol{s}',\Delta\boldsymbol{x}')\right\rangle\right|^2\\&=\left|\left\langle \eta\alpha(\boldsymbol{x},\boldsymbol{s},\Delta\boldsymbol{x})\middle|(\boldsymbol{s},\Delta\boldsymbol{x}\,|\,\boldsymbol{x})\right\rangle\right|^4\left|\left\langle (\boldsymbol{s}',\Delta\boldsymbol{x}'\,|\,\boldsymbol{x})\middle|\eta\alpha(\boldsymbol{x},\boldsymbol{s},\Delta\boldsymbol{x})\right\rangle\right|^2\end{aligned}. \tag{27}$$

*4.3 Security of Y00 under KPA on Secret Key: in case of Exact Signal Detections for Eve*

Although it is impossible for Eve to obtain the correct signal sequence without any errors because of quantum noise in Y00 protocol, it is worth to consider an imaginary case where Eve could detect signals without any errors to compare Y00 protocol with conventional stream ciphers in Sec.3.

The situation where Eve could detect signals without any errors is that, from the Born rule,

$$\left|\left\langle (\boldsymbol{s},\Delta\boldsymbol{x}\,|\,\boldsymbol{x})\middle|\eta\alpha(\boldsymbol{x},\boldsymbol{s}'\neq\boldsymbol{s},\Delta\boldsymbol{x}'\neq\Delta\boldsymbol{x})\right\rangle\right|^2=0\,. \tag{28}$$

Eq.(28) also implies from Eq.(21) that

$$\left\langle \eta\alpha(\boldsymbol{x},\boldsymbol{s}',\Delta\boldsymbol{x}')\middle|\eta\alpha(\boldsymbol{x},\boldsymbol{s}',\Delta\boldsymbol{x}')\right\rangle=1=\left|\left\langle \eta\alpha(\boldsymbol{x},\boldsymbol{s},\Delta\boldsymbol{x})\middle|(\boldsymbol{s},\Delta\boldsymbol{x}\,|\,\boldsymbol{x})\right\rangle\right|^2. \tag{29}$$

Then, from the left-hand side of Eq.(22),

$$-\operatorname{tr}\Gamma=\sum\nolimits_{(\boldsymbol{s},\Delta\boldsymbol{x})\in(\boldsymbol{S},\Delta\boldsymbol{X})}\Pr(\boldsymbol{s},\Delta\boldsymbol{x})\left|\left\langle \eta\alpha(\boldsymbol{x},\boldsymbol{s},\Delta\boldsymbol{x})\middle|(\boldsymbol{s},\Delta\boldsymbol{x}\,|\,\boldsymbol{x})\right\rangle\right|^2=1\,. \tag{30}$$

Therefore, through one period of ($\boldsymbol{s}$, $\Delta\boldsymbol{x}$), that is $T_{\mathrm{LCM}}$, Eve would obtain the correct ($\boldsymbol{s}$, $\Delta\boldsymbol{x}$) with a probability of 1. Then the situation is the same as conventional stream ciphers. Therefore, the effect of unavoidable quantum noise in Eq.(28) as a non-zero factor should play an important role in Y00 protocol.

*4.4 Security of Y00 under KPA on Secret Key: in case of Erroneous Signal Detections for Eve*

Unless Eve's detections are error-free expressed by Eq.(28), from Eq.(21),

$$\left|\left\langle \eta\alpha(\boldsymbol{x},\boldsymbol{s},\Delta\boldsymbol{x})\middle|(\boldsymbol{s},\Delta\boldsymbol{x}\mid\boldsymbol{x})\right\rangle\right|^{2} < 1 . \tag{31}$$

Therefore, Eq.(24) satisfies the following inequality as well.

$$\max\left[-\mathrm{tr}\,\Gamma\right] = \frac{\sum_{(\boldsymbol{s},\Delta\boldsymbol{x})\in(\boldsymbol{S},\Delta\boldsymbol{X})}\left|\left\langle \eta\alpha(\boldsymbol{x},\boldsymbol{s},\Delta\boldsymbol{x})\middle|(\boldsymbol{s},\Delta\boldsymbol{x}\mid\boldsymbol{x})\right\rangle\right|^{4}}{\sum_{(\boldsymbol{s},\Delta\boldsymbol{x})\in(\boldsymbol{S},\Delta\boldsymbol{X})}\left|\left\langle \eta\alpha(\boldsymbol{x},\boldsymbol{s},\Delta\boldsymbol{x})\middle|(\boldsymbol{s},\Delta\boldsymbol{x}\mid\boldsymbol{x})\right\rangle\right|^{2}} < 1 . \tag{32}$$

Even if Pr($\boldsymbol{s}$, $\Delta\boldsymbol{x}$) is uniform, that is Pr($\boldsymbol{s}$, $\Delta\boldsymbol{x}$) = $2^{-|\boldsymbol{K}|-|\Delta\boldsymbol{K}|}$, since Eve has to make the success probability in measurement larger than the failure probability,

$$\left|\left\langle \eta\alpha(\boldsymbol{x},\boldsymbol{s},\Delta\boldsymbol{x})\middle|(\boldsymbol{s},\Delta\boldsymbol{x}\mid\boldsymbol{x})\right\rangle\right|^{2} \geq \left|\left\langle \eta\alpha(\boldsymbol{x},\boldsymbol{s},\Delta\boldsymbol{x})\middle|(\boldsymbol{s}',\Delta\boldsymbol{x}'\mid\boldsymbol{x})\right\rangle\right|^{2} . \tag{33}$$

Then from Eqs.(21, 22),

$$\begin{aligned} -\mathrm{tr}\,\Gamma &= \sum_{(\boldsymbol{s},\Delta\boldsymbol{x})\in(\boldsymbol{S},\Delta\boldsymbol{X})}\Pr(\boldsymbol{s},\Delta\boldsymbol{x})\left|\left\langle \eta\alpha(\boldsymbol{x},\boldsymbol{s},\Delta\boldsymbol{x})\middle|(\boldsymbol{s},\Delta\boldsymbol{x}\mid\boldsymbol{x})\right\rangle\right|^{2} \\ &= 2^{-|\boldsymbol{K}|-|\Delta\boldsymbol{K}|}\sum_{(\boldsymbol{s},\Delta\boldsymbol{x})\in(\boldsymbol{S},\Delta\boldsymbol{X})}\left|\left\langle \eta\alpha(\boldsymbol{x},\boldsymbol{s},\Delta\boldsymbol{x})\middle|(\boldsymbol{s},\Delta\boldsymbol{x}\mid\boldsymbol{x})\right\rangle\right|^{2} \geq 2^{-|\boldsymbol{K}|-|\Delta\boldsymbol{K}|} \end{aligned} . \tag{34}$$

Therefore, Eve has an advantage in obtaining the correct ($\boldsymbol{s}$, $\Delta\boldsymbol{x}$) compared to pure-guessing. Thus, even Eve launches KPA using Quantum Multiple Hypotheses Testing Theory during a least common multiple of the periods of two PRNGs, she cannot pin-down the keys deterministically, far different from conventional stream ciphers. The problem is how long Y00 protocol stays secure.

## 5 Security of Y00 Protocol under Unlimitedly Long KPA

This section describes the security of Y00 protocol under unlimitedly long KPA so that Eve guesses the most likely key by Bayes Criterion [29].

### *5.1 Y00 Protocol under Unlimitedly Long KPA*

Since ($\boldsymbol{s}$, $\Delta\boldsymbol{x}$) is pseudo-random of a period of $T_{\mathrm{LCM}}$ while the plaintext $\boldsymbol{x}$ is supposed not to repeat, Eve can statistically confirm the most likely ($\boldsymbol{s}$, $\Delta\boldsymbol{x}$) during $N\cdot T_{\mathrm{LCM}}$ periods as shown in Table 1.

**Table 1** A time table of a set of variables ($\boldsymbol{m}$, $\boldsymbol{s}$, $\Delta\boldsymbol{x}$, $\boldsymbol{x}$).

| $t$ | 0 | 1 | … | $T_{\text{LCM}}-1$ | $T_{\text{LCM}}$ | … | $2T_{\text{LCM}}-1$ | $2T_{\text{LCM}}$ | … | $3T_{\text{LCM}}-1$ | $3T_{\text{LCM}}$ | … |
|---|---|---|---|---|---|---|---|---|---|---|---|---|
| $\boldsymbol{m}$ | $m(0)$ | $m(1)$ | … | $m(T_{\text{LCM}}-1)$ | $m(T_{\text{LCM}})$ | … | $m(2T_{\text{LCM}}-1)$ | $m(2T_{\text{LCM}}+1)$ | … | $m(3T_{\text{LCM}}-1)$ | $m(3T_{\text{LCM}})$ | … |
| $\boldsymbol{s}$ | $\boldsymbol{s}(0)$ | $\boldsymbol{s}(1)$ | … | $\boldsymbol{s}(T_{\text{LCM}}-1)$ | $\boldsymbol{s}(0)$ | … | $\boldsymbol{s}(T_{\text{LCM}}-1)$ | $\boldsymbol{s}(0)$ | … | $\boldsymbol{s}(T_{\text{LCM}}-1)$ | $\boldsymbol{s}(0)$ | … |
| $\Delta\boldsymbol{x}$ | $\Delta x(0)$ | $\Delta x(1)$ | … | $\Delta x(T_{\text{LCM}}-1)$ | $\Delta x(0)$ | … | $\Delta x(T_{\text{LCM}}-1)$ | $\Delta x(0)$ | … | $\Delta x(T_{\text{LCM}}-1)$ | $\Delta x(0)$ | … |
| $\boldsymbol{x}$ | $x(0)$ | $x(1)$ | … | $x(T_{\text{LCM}}-1)$ | $x(T_{\text{LCM}})$ | … | $x(2T_{\text{LCM}}-1)$ | $x(2T_{\text{LCM}})$ | … | $x(3T_{\text{LCM}}-1)$ | $x(3T_{\text{LCM}})$ | … |

At the $n$-th period of $n \in \{1, 2, 3, \ldots, N\}$, Eve measures coherent states $\left|\eta\alpha(\boldsymbol{x}_n, \boldsymbol{s}, \Delta\boldsymbol{x})\right\rangle$ with a set of operators denoted as $\{E(\boldsymbol{s}, \Delta\boldsymbol{x} \mid \boldsymbol{x}_n)\}$ based on known plaintext $\boldsymbol{x}_n$,

$$\left[W(\boldsymbol{x}_n, \boldsymbol{s}, \Delta\boldsymbol{x}) - \Gamma\right] E(\boldsymbol{s}, \Delta\boldsymbol{x} \mid \boldsymbol{x}_n) = E(\boldsymbol{s}, \Delta\boldsymbol{x} \mid \boldsymbol{x}_n)\left[W(\boldsymbol{x}_n, \boldsymbol{s}, \Delta\boldsymbol{x}) - \Gamma\right] = 0\,. \tag{35}$$

$$E(\boldsymbol{s}, \Delta\boldsymbol{x} \mid \boldsymbol{x}_n)\left[W(\boldsymbol{x}_n, \boldsymbol{s}', \Delta\boldsymbol{x}') - W(\boldsymbol{x}_n, \boldsymbol{s}, \Delta\boldsymbol{x})\right] E(\boldsymbol{s}', \Delta\boldsymbol{x}' \mid \boldsymbol{x}_n) = \boldsymbol{0}\,. \tag{36}$$

$$W(\boldsymbol{x}_n, \boldsymbol{s}, \Delta\boldsymbol{x}) - \Gamma_n \geq 0\,. \tag{37}$$

$$\Gamma_n := \sum_{(\boldsymbol{s},\Delta\boldsymbol{x})\in(\boldsymbol{S},\Delta\boldsymbol{X})} E(\boldsymbol{s}, \Delta\boldsymbol{x} \mid \boldsymbol{x}_n) W(\boldsymbol{x}_n, \boldsymbol{s}, \Delta\boldsymbol{x}) = \sum_{(\boldsymbol{s},\Delta\boldsymbol{x})\in(\boldsymbol{S},\Delta\boldsymbol{X})} W(\boldsymbol{x}_n, \boldsymbol{s}, \Delta\boldsymbol{x}) E(\boldsymbol{s}, \Delta\boldsymbol{x} \mid \boldsymbol{x}_n)\,. \tag{38}$$

Since Eve just performs $2^{|\boldsymbol{K}|+|\Delta\boldsymbol{K}|}$-ary quantum hypotheses testing to obtain ($\boldsymbol{s}$, $\Delta\boldsymbol{x}$) based on known $\boldsymbol{x}_n$, the results are independent of $n$. Therefore, from the Born rule, define as follows.

$$\left|\left\langle(\boldsymbol{s}, \Delta\boldsymbol{x} \mid \boldsymbol{x}_n)\right|\eta\alpha(\boldsymbol{x}_n, \boldsymbol{s}', \Delta\boldsymbol{x}')\right\rangle\right|^2 := \Pr(\boldsymbol{s}, \Delta\boldsymbol{x} \mid \boldsymbol{x}_0, \boldsymbol{s}', \Delta\boldsymbol{x}')\,. \tag{39}$$

$$\Gamma_0 := \Gamma_n\,. \tag{40}$$

Suppose that Eve has obtained $n(\boldsymbol{s}, \Delta\boldsymbol{x})$ times of her measurement result ($\boldsymbol{s}$, $\Delta\boldsymbol{x}$) during $N{\cdot}T_{\text{LCM}}$ periods, then such a probability is

$$\Pr(n(\boldsymbol{s}, \Delta\boldsymbol{x}) \mid \boldsymbol{x}_0, \boldsymbol{s}, \Delta\boldsymbol{x}) := {}_N C_{n(\boldsymbol{s},\Delta\boldsymbol{x})} \Pr(\boldsymbol{s}, \Delta\boldsymbol{x} \mid \boldsymbol{x}_0, \boldsymbol{s}, \Delta\boldsymbol{x})^{n(\boldsymbol{s},\Delta\boldsymbol{x})}\left[1 - \Pr(\boldsymbol{s}, \Delta\boldsymbol{x} \mid \boldsymbol{x}_0, \boldsymbol{s}, \Delta\boldsymbol{x})\right]^{N-n(\boldsymbol{s},\Delta\boldsymbol{x})}. \tag{41}$$

At the boundary where Eve makes a wrong decision, Bayes Criterion requests Eq.(42),

$$\Pr(\boldsymbol{s}, \Delta\boldsymbol{x})\Pr(n_{\text{Th}} \mid \boldsymbol{x}_0, \boldsymbol{s}, \Delta\boldsymbol{x}) = \Pr(\boldsymbol{s}', \Delta\boldsymbol{x}')\Pr(n_{\text{Th}} \mid \boldsymbol{x}_0, \boldsymbol{s}', \Delta\boldsymbol{x}')\,. \tag{42}$$

Two nearest probability distributions give the following boundary conditions.

$$\Pr(n_{\text{Th}} \mid \boldsymbol{x}_0, \boldsymbol{s}, \Delta\boldsymbol{x}) = {}_N C_{n_{\text{Th}}} \Pr(\boldsymbol{s}, \Delta\boldsymbol{x} \mid \boldsymbol{x}_0, \boldsymbol{s}, \Delta\boldsymbol{x})^{n_{\text{Th}}}\left[1 - \Pr(\boldsymbol{s}, \Delta\boldsymbol{x} \mid \boldsymbol{x}_0, \boldsymbol{s}, \Delta\boldsymbol{x})\right]^{N-n_{\text{Th}}}\,. \tag{43}$$

$$\Pr\left(n_{\mathrm{Th}} \mid \boldsymbol{x}_0, \boldsymbol{s}', \Delta\boldsymbol{x}'\right) = {}_N C_{n_{\mathrm{Th}}} \Pr\left(\boldsymbol{s}, \Delta\boldsymbol{x} \mid \boldsymbol{x}_0, \boldsymbol{s}', \Delta\boldsymbol{x}'\right)^{n_{\mathrm{Th}}} \left[1 - \Pr\left(\boldsymbol{s}, \Delta\boldsymbol{x} \mid \boldsymbol{x}_0, \boldsymbol{s}', \Delta\boldsymbol{x}'\right)\right]^{N-n_{\mathrm{Th}}} . \tag{44}$$

Substituting Eqs.(43, 44) into Eq.(42), $n_{\mathrm{Th}}$ is given in Eq.(46).

$$\begin{aligned} &\log_2 \frac{\Pr\left(\boldsymbol{s}, \Delta\boldsymbol{x}\right)\Pr\left(n_{\mathrm{Th}} \mid \boldsymbol{x}_0, \boldsymbol{s}, \Delta\boldsymbol{x}\right)}{\Pr\left(\boldsymbol{s}', \Delta\boldsymbol{x}'\right)\Pr\left(n_{\mathrm{Th}} \mid \boldsymbol{x}_0, \boldsymbol{s}', \Delta\boldsymbol{x}'\right)} = 0 \\ &= \log_2 \frac{\Pr\left(\boldsymbol{s}, \Delta\boldsymbol{x}\right)}{\Pr\left(\boldsymbol{s}', \Delta\boldsymbol{x}'\right)} + n_{\mathrm{Th}} \log_2 \frac{\Pr\left(\boldsymbol{s}, \Delta\boldsymbol{x} \mid \boldsymbol{x}_0, \boldsymbol{s}, \Delta\boldsymbol{x}\right)}{\Pr\left(\boldsymbol{s}, \Delta\boldsymbol{x} \mid \boldsymbol{x}_0, \boldsymbol{s}', \Delta\boldsymbol{x}'\right)} + \left(N - n_{\mathrm{Th}}\right)\log_2 \frac{1 - \Pr\left(\boldsymbol{s}, \Delta\boldsymbol{x} \mid \boldsymbol{x}_0, \boldsymbol{s}, \Delta\boldsymbol{x}\right)}{1 - \Pr\left(\boldsymbol{s}, \Delta\boldsymbol{x} \mid \boldsymbol{x}_0, \boldsymbol{s}', \Delta\boldsymbol{x}'\right)} \end{aligned} . \tag{45}$$

$$n_{\mathrm{Th}} = \frac{N \log_2 \dfrac{1 - \Pr\left(\boldsymbol{s}, \Delta\boldsymbol{x} \mid \boldsymbol{x}_0, \boldsymbol{s}', \Delta\boldsymbol{x}'\right)}{1 - \Pr\left(\boldsymbol{s}, \Delta\boldsymbol{x} \mid \boldsymbol{x}_0, \boldsymbol{s}, \Delta\boldsymbol{x}\right)} + \log_2 \dfrac{\Pr\left(\boldsymbol{s}', \Delta\boldsymbol{x}'\right)}{\Pr\left(\boldsymbol{s}, \Delta\boldsymbol{x}\right)}}{\log_2 \dfrac{\Pr\left(\boldsymbol{s}, \Delta\boldsymbol{x} \mid \boldsymbol{x}_0, \boldsymbol{s}, \Delta\boldsymbol{x}\right)\left[1 - \Pr\left(\boldsymbol{s}, \Delta\boldsymbol{x} \mid \boldsymbol{x}_0, \boldsymbol{s}', \Delta\boldsymbol{x}'\right)\right]}{\Pr\left(\boldsymbol{s}, \Delta\boldsymbol{x} \mid \boldsymbol{x}_0, \boldsymbol{s}', \Delta\boldsymbol{x}'\right)\left[1 - \Pr\left(\boldsymbol{s}, \Delta\boldsymbol{x} \mid \boldsymbol{x}_0, \boldsymbol{s}, \Delta\boldsymbol{x}\right)\right]}} . \tag{46}$$

This situation is depicted in Fig.1.

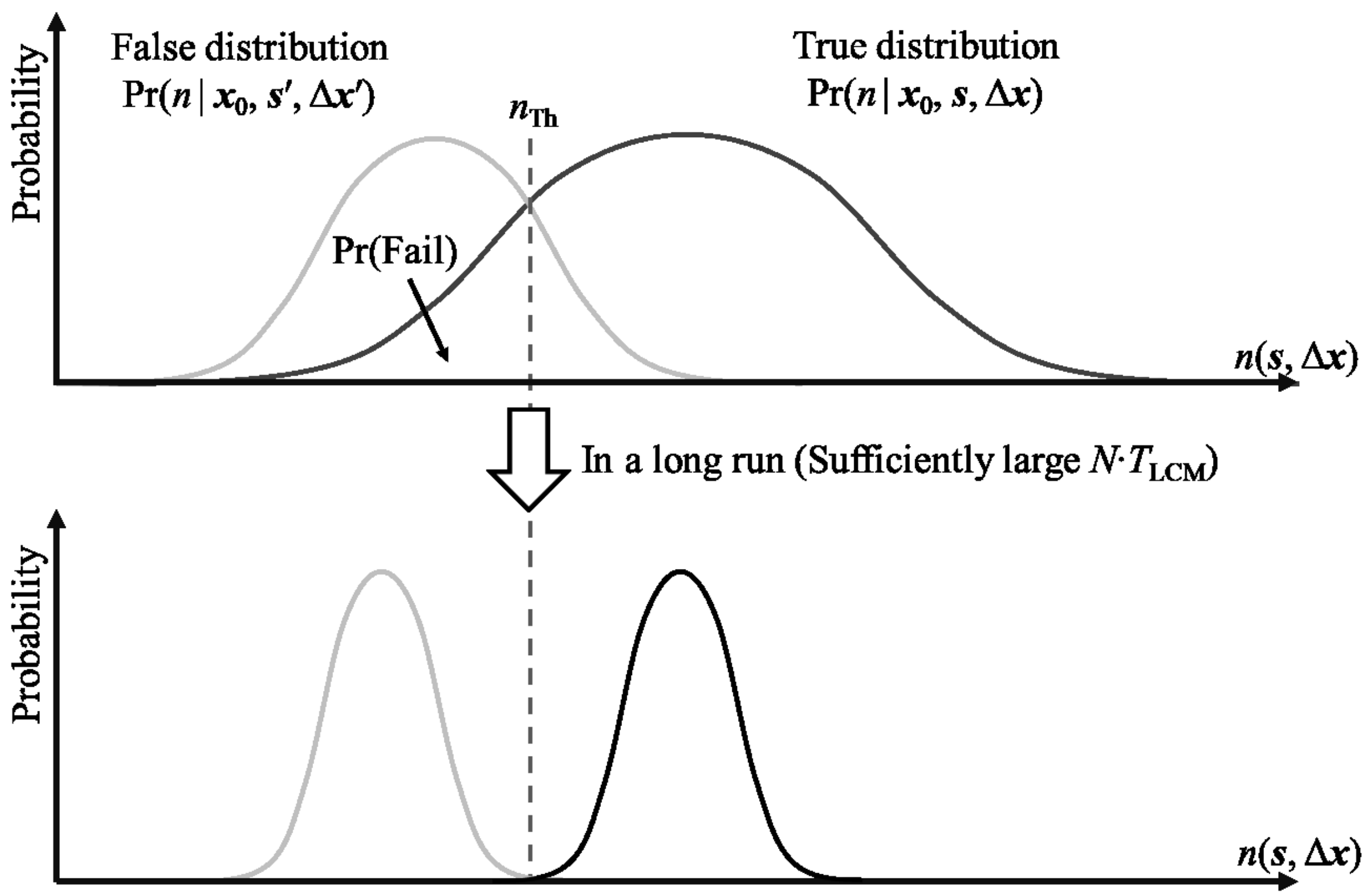

**Fig.1** Schematic view of how the seucrity of Y00 system is evaluated by Eve's failure probability.

Since there are $2^{|\boldsymbol{K}|+|\Delta\boldsymbol{K}|}$ patterns of possible probability distributions, and only one is for the correct ($\boldsymbol{s}$, $\Delta\boldsymbol{x}$). Therefore, maximizing $n_{\mathrm{Th}}$ by all wrong sets of ($\boldsymbol{s}'$, $\Delta\boldsymbol{x}'$) and defining it as max $n_{\mathrm{Th}}$,

$$\Pr(\mathrm{Fail}) = \sum_{n(\boldsymbol{s},\Delta\boldsymbol{x})=0}^{\max n_{\mathrm{Th}}} {}_{N}C_{n(\boldsymbol{s},\Delta\boldsymbol{x})} \Pr(\boldsymbol{s},\Delta\boldsymbol{x} \mid \boldsymbol{x}_0, \boldsymbol{s}, \Delta\boldsymbol{x})^{n(\boldsymbol{s},\Delta\boldsymbol{x})} \left[1 - \Pr(\boldsymbol{s},\Delta\boldsymbol{x} \mid \boldsymbol{x}_0, \boldsymbol{s}, \Delta\boldsymbol{x})\right]^{N-n(\boldsymbol{s},\Delta\boldsymbol{x})} . \quad (47)$$

Thus, Eve's success probability in obtaining the correct ($\boldsymbol{s}$, $\Delta\boldsymbol{x}$) corresponding to the shared secret keys ($\boldsymbol{k}$, $\Delta\boldsymbol{k}$) in the Y00 system is Pr(Success) = 1 – Pr(Fail).

To perform numerical simulation, all conditional probabilities $\{\Pr(\boldsymbol{s}, \Delta\boldsymbol{x} \mid \boldsymbol{x}_0, \boldsymbol{s}', \Delta\boldsymbol{x}')\}$ defined in Eq.(39) have to be determined. However, those parameters are dependent on implementations of Y00 systems including key-expansion algorithms. Therefore, this study gives numerical examples as follows. Assume initial key lengths are $|\boldsymbol{K}| = |\Delta\boldsymbol{K}| = 128$ bits and

$$\Pr(\boldsymbol{s},\Delta\boldsymbol{x} \mid \boldsymbol{x}_0, \boldsymbol{s}, \Delta\boldsymbol{x}) = \Pr(\boldsymbol{x}',\Delta\boldsymbol{x}' \mid \boldsymbol{x}_0, \boldsymbol{x}', \Delta\boldsymbol{x}') := p . \quad (48)$$

$$\Pr\left({}^{\forall}\boldsymbol{s}' \neq \boldsymbol{s}, {}^{\forall}\Delta\boldsymbol{x}' \neq \Delta\boldsymbol{x} \mid \boldsymbol{x}_0, \boldsymbol{s}, \Delta\boldsymbol{x}\right) = \Pr\left({}^{\forall}\boldsymbol{s} \neq \boldsymbol{s}', {}^{\forall}\Delta\boldsymbol{x} \neq \Delta\boldsymbol{x}' \mid \boldsymbol{x}_0, \boldsymbol{s}', \Delta\boldsymbol{x}'\right) = (1-p)/\left(2^{2\times128}-1\right) . \quad (49)$$

$$\Pr\left({}^{\forall}\boldsymbol{s}, {}^{\forall}\Delta\boldsymbol{x}\right) = 2^{-2\times128} . \quad (50)$$

The numerical simulation result with the above situation is shown in Fig.2.

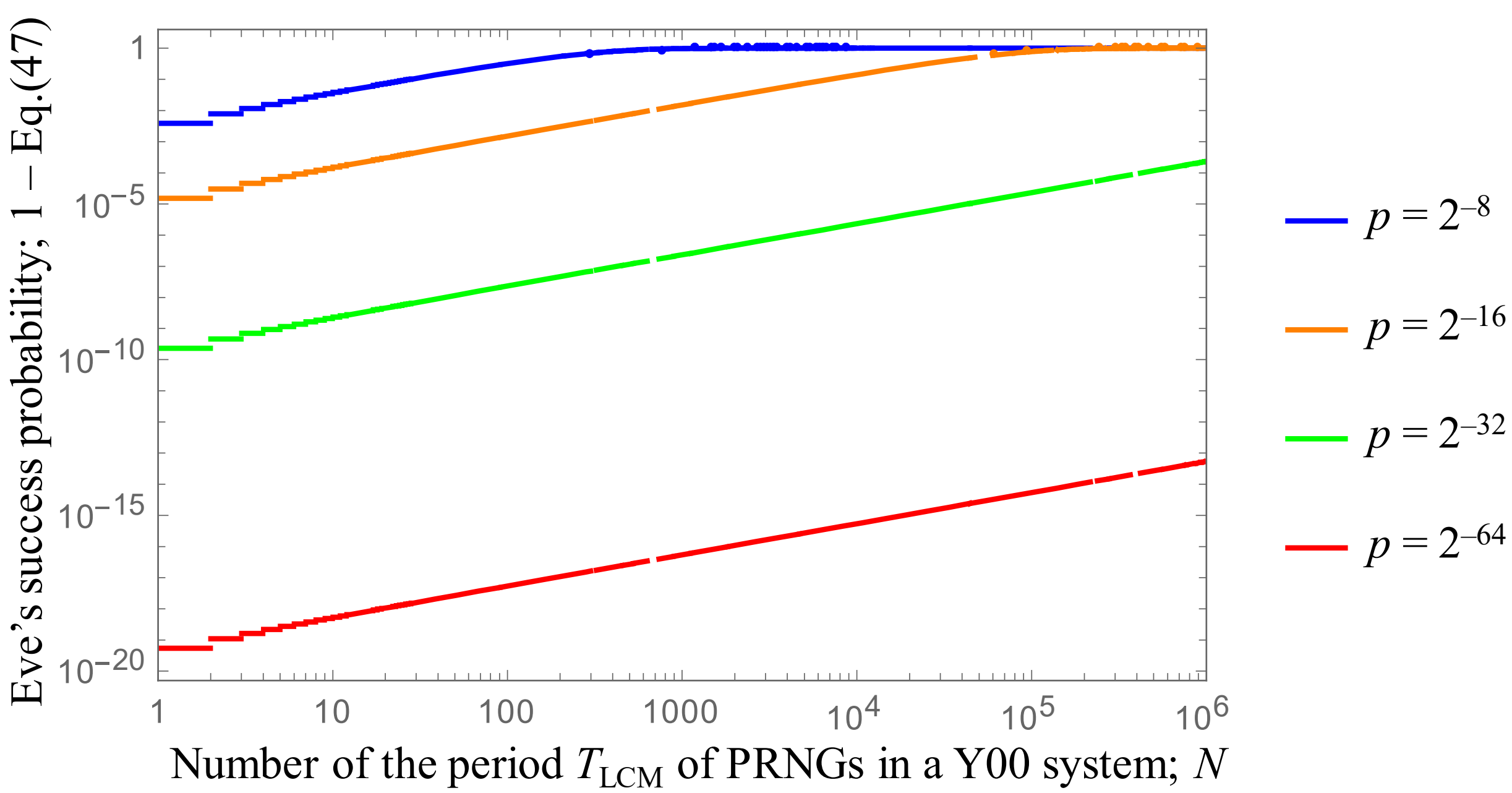

**Fig.2** Schematic view of how the seucrity of Y00 system is evaluated by Eve's failure probability.

As Eve's success probability of obtaining the correct ($\boldsymbol{s}$, $\Delta\boldsymbol{x}$) in one $T_{\mathrm{LCM}}$ smaller, Eve needs a larger number of $N$ which is a repetition number of the period $T_{\mathrm{LCM}}$ of the two PRNGs. However, note that even with $p = 2^{-16}$, Eve needs $N = 10^4$ periods to pin-down the correct ($\boldsymbol{k}$, $\Delta\boldsymbol{k}$) with a probability of almost 1. When $p = 2^{-64}$, even $N = 10^7$ periods are not enough for Eve, only allowing her to guess the correct ($\boldsymbol{k}$, $\Delta\boldsymbol{k}$) with a probability of about $10^{-13}$. Therefore it was shown that Y00 protocol can go beyond the Shannon Limit of cryptography [30].

While Eve's success probability is small enough, Alice has to send fresh keys to Bob to continue secure quantum communications. In this case, Eve's probability of obtaining the fresh keys has to be evaluated by COA with probabilistically known ($\boldsymbol{k}$, $\Delta\boldsymbol{k}$), that is,

$$\Pr\left(\boldsymbol{k}_{\mathrm{new}}, \Delta\boldsymbol{k}_{\mathrm{new}}\right) = \sum_{(\boldsymbol{s},\Delta\boldsymbol{x})\in(\boldsymbol{S},\Delta\boldsymbol{X})} \Pr\left(\boldsymbol{s}, \Delta\boldsymbol{x}\right) \Pr\left(\boldsymbol{k}_{\mathrm{new}}, \Delta\boldsymbol{k}_{\mathrm{new}} \mid \boldsymbol{s}, \Delta\boldsymbol{x}\right). \tag{51}$$

## 6 Results and Discussions: Assumptions used in Security Analysis and Future Works

In this work, it was shown that even a Y00 system with a few-hundreds bits of shared keys can be secure far longer than the period of its Pseudo-Random Number Generators, such as $10^7$ periods if its implementations are well-designed. The security analyses in this study applied the following assumptions.

1. Irregular-Mapping [3] makes quantum noise cover all bits in the chopped key stream equally.
2. Irregular-Mapping is fixed and known to Eve during her eavesdropping.

Hence, there are still necessities of further studies to give mathematically more rigorous analyses when the above assumptions are not satisfied. Also, the security of fresh keys is not given in this study yet. Hence it has to be analyzed in the next work.

**Simulation code for Fig.2 on Mathematica 11.3**

```
ps[p_] := p;

pf[p_] := (1 - ps[p])/(2^(128*2)-1);

nth[n_, p_] := n*Log[2, (1 - pf[p])/(1 - ps[p])]/Log[2, ps[p]*(1 - pf[p]) / pf[p] / (1 - ps[p])];

prob[n_,p_] := 1 - CDF[BinomialDistribution[n, ps[p]], Floor[nth[n, p]]];

Show[ Table[ LogLogPlot[ prob[ Floor[m], 2^(-b)], {m, 1,10^6},

    PlotRange -> {{1, 10^6}, {5*10^(-21), 4}}, Frame -> True,

    PlotLegends -> {Switch[b, 8, "p=2^-8", 16, "p=2^-16", 32, "p=2^-32", 64, "p=2^-64"]},

    PlotStyle->{Switch[b, 8, Blue, 16, Orange, 32, Green, 64, Red]}], {b, {8, 16, 32, 64}}]]
```